\documentclass{ws-p8-50x6-00}

\newcommand{\insertfig}[2]{\mbox{\epsfxsize=#1cm \epsfbox{#2.eps}}}

\begin{document}


\thispagestyle{empty}

\begin{flushright}
DOE/ER/40762-253 \\
UMD-PP\#02-044 \\
\end{flushright}

\vspace{5mm}

\centerline{\large \bf QCD evolution equations}

\vspace{10mm}

\centerline{\bf A.V. Belitsky}

\vspace{5mm}

\centerline{\it Department of Physics}
\centerline{\it University of Maryland at College Park}
\centerline{\it College Park, MD 20742-4111, USA}

\vspace{15mm}

\centerline{\bf Abstract}

\vspace{5mm}

We discuss QCD evolution equations for two and three particle correlation
functions of quarks and gluon fields in a hadron which describe development
of the momentum distribution of a parton system with a change of the wave
length of a probe which resolves it. We show in a general case of
two-particle correlators how the four-dimensional conformal algebra and the
known pattern of conformal symmetry breaking in QCD can be used to solve
the complicated mixing problem of local operators under renormalization and
compute economically anomalous dimensions of quark and gluon composite
operators. An extension of QCD to ${\cal N} = 1$ super Yang-Mills theory
and use of superconformal anomalies arising after quantization allows to
derive non-trivial relations between the anomalous dimensions. For
three-parton systems the conformal symmetry alone is not enough to solve the
three-particle problem. We show that in milticolor limit of QCD there arises
an extra conserved charge describing the solitonic motion of the system of
particles. The problem admits a one-to-one correspondence with certain spin
chain models which are exactly solvable.

\vspace{15mm}

\centerline{\it Talk given at the}
\centerline{\it Workshop on The Phenomenology of Large $N_c$ QCD}
\centerline{\it Tempe, Arizona, January 9-11, 2002}

\setcounter{page}{0}


\title{QCD evolution equations}

\author{A.V. Belitsky}

\address{Department of Physics \\
University of Maryland at College Park \\
College Park, MD 20742-4111, USA}

\maketitle

\abstracts{We discuss QCD evolution equations for two and three particle
correlation functions of quarks and gluon fields in a hadron which describe
development of the momentum distribution of a parton system with a change
of the wave length of a probe which resolves it. We demonstrate how broken
space-time and hidden QCD symmetries serve to understand the structure and
ultimately solve these equations.}

\section{Unraveling layers of matter: from atoms to partons}

A wisdom goes back to ancient Greeks who philosophized that the matter
consists of tiny particles --- atoms. However, the atomic structure remained
a puzzle till the beginning of the 20th century when the radioactivity was
discovered and used by Rutherford in his seminal experiments on large angle
scattering of $\alpha$ particles off atoms which suggested that the atom bears
a localized core --- the nucleus. On the other hand, electron beams were found
to pass through atoms with no or very little deflection forcing Lenard to
hypothesize that atoms have wide empty spaces inside. $\alpha$ particles having
much larger wave length comparable to the nucleus size scattered more frequently
with a low intensity beams than their `cousins' $\beta$ particles having much
smaller wave length and which, for available luminosity, had a very low cross
section. Similar experiments but rather with light sources or neutrons are
exploited nowadays to study the crystal structure. If one puts a crystal in
front of a source of visible light, the object just leaves a shadow on a screen
behind it and one does not see elementary building blocks which form it. Obviously,
the visible light, having the wave length $\lambda_{\rm light} \sim 0.4 - 0.7 \
\mu{\rm m}$ cannot do the job and resolve the internal structure of a crystal. The
size of an individual atom, say hydrogen, is of order $r_{\rm atom} \sim \left(
\alpha_{\rm em} m_e \right)^{-1} \sim \left( 10 \ {\rm KeV} \right)^{- 1}$.
Therefore, to `see' atoms in crystals one has to have photons with a comparable
or smaller wave length $\lambda_\gamma \leq r_{\rm atom}$, or equivalently, of
energy $E_\gamma \geq r^{-1}_{\rm atom}$. To do this kind of `nano-photography'
one need a beam of X-rays. To go beyond and look into the structure of atoms
one needs even smaller wave lengths.

\begin{figure}[t]
\unitlength1mm
\begin{center}
\mbox{
\begin{picture}(0,24)(58,0)
\put(0,0){\insertfig{5}{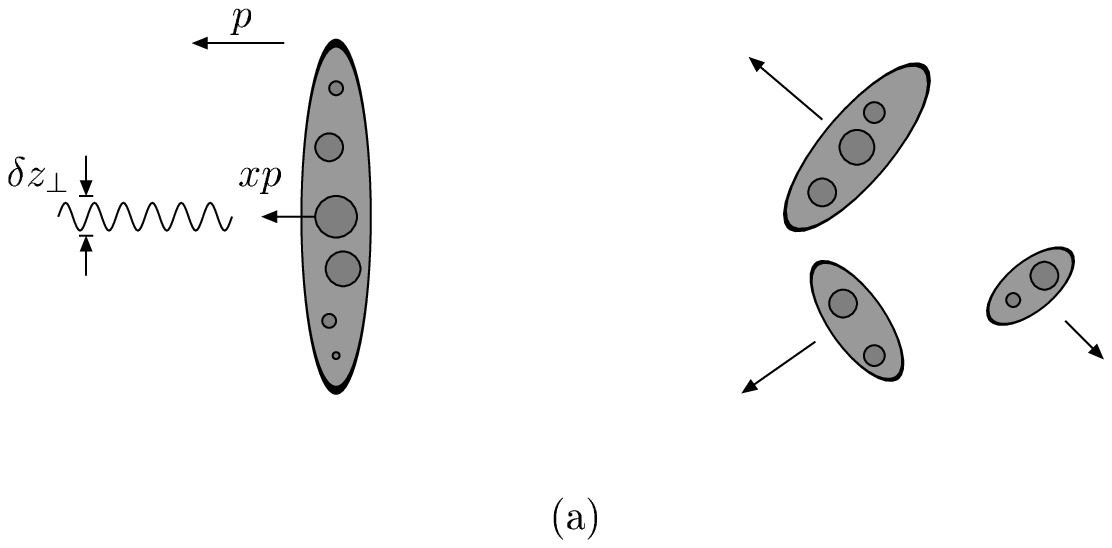}}
\put(65,0){\insertfig{5}{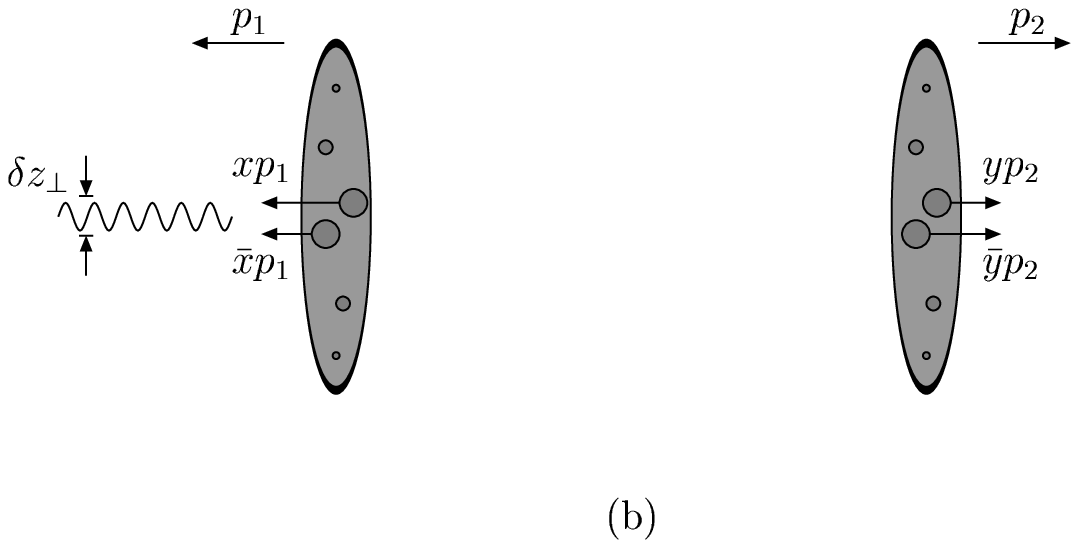}}
\end{picture}
}
\end{center}
\caption{\label{PartonPicture} Partonic picture of inclusive (a) and
exclusive (b) scattering of a $\gamma^\ast$-probe from the nucleon.}
\end{figure}

After the discovery of nucleus' building blocks --- nucleons, i.e.,
protons and neutrons, the attention has been shifted to the study of these
`elementary' particles. However, their elementarity has been questioned
starting from Stern's measurements in 1933 that demonstrated a large
proton's magnetic moment. Hofstadter's experiments with elastic electron
scattering off nucleons, $e N \to e' N'$, and measurement of their charge
distribution revealed that the nucleon has a spatial extent of order 1 fm.
To probe femtometer scales one has resort to scattering experiments with
high energy beams. Inelastic lepton-nucleon scattering experiments,
$e N \to e' X$, conducted at SLAC fulfilled this goal and led finally to
the discovery of a new layer of matter by observing events with the transfer
of a large momentum from the electron to the proton. If the latter would be
a hard ball or a diffuse distribution of matter, such kind of scattering
would be improbable. Rather it was explained by conjecturing the existence
of point-like constituents inside the nucleon which absorb a highly virtual
$\gamma^\ast$-quantum emitted by leptons, see Fig.\ \ref{PartonPicture} (a).
Analogous representations holds for the elastic scattering alluded to above,
Fig.\ \ref{PartonPicture} (b). These partons are nothing else than QCD quark
and gluons described by a rigid $SU (3)$ gauge field theory.

\section{Hard scattering and QCD}

A QCD scattering amplitude is given by a successive sequence of interactions
and free propagations through rife intermediate states created by QCD field
operators. At $n$th order in the coupling $g = \sqrt{4 \pi \alpha_s}$ it reads
in the time-ordered form
\begin{eqnarray*}
\int\limits_{- \infty}^{\infty} d t_n
\int\limits_{- \infty}^{t_n} d t_{n - 1}
\dots
\int\limits_{- \infty}^{t_2} d t_1 \
G_{n, n - 1} (t_n) \, V_{n - 1} \,
G_{n - 1, n - 2} (t_{n - 1}) \,
\dots
V_{1} \, G_{1, 0} (t_1) \, .
\end{eqnarray*}
The summation in repeated indices runs over states which are eigenfunctions of
free equations of motion. $V_{n, m} = \langle n | V | m \rangle$ is a matrix
element of a perturbation, --- the interaction potential, --- and the free
particle propagation is expressed in terms of a time-underintegrated free action,
$S$, which defines a Hamiltonian of the noninteracting system,
\begin{eqnarray*}
G_{n, m} (t) = \exp \left\{ i t \, \left( S_n - S_m \right) \right\} \, .
\end{eqnarray*}
At this point, one immediately concludes that there is no sensitivity to
long-time scales as long as the total phase, the free particle action is time
dependent \cite{Ste00}. In this case, cancellations of long-time contributions
occur due to oscillatory nature of the integrand. However, if the phase is
stationary, which corresponds to a stationary classical action, particles
travel along their classical trajectories and the amplitude becomes sensitive
to large-time dynamics. Since at large space-time scales $\alpha_s$ is large,
one cannot evaluate the amplitude in a perturbative expansion and effects of
quark confinement are relevant. In the opposite situation, which goes under
the name of infrared safe, an observable can be computed as a power series
in QCD coupling alone. These include the total inclusive $\sigma_{\rm tot}$
and jet $\sigma_{\rm jet}$ cross sections in electron-positron annihilation,
etc. Indeed in $\sigma_{\rm tot} \sim \langle 0 | T \{ j^\dagger_\mu j_\mu \}
| 0 \rangle$, once a pair of quark and antiquark is created by the source
$j_\mu$, they travel back-to-back with a speed of light and cannot reassemble
back into a physical state absorbed by $j^\dagger_\mu$. Since there are no
classical trajectories which allow this processes, the observable is not
sensitive to infrared physics according to the criterion spelled out above.

A hard scattering cross section having at least one hadron in the initial
state cannot be infrared safe since in a preparation of the asymptotic
hadron state partons which form it had interacted strongly with each
other for a long time in a bound state. Obviously, the wave functions of
quarks in a bound state differ from what one would have if they were free.
So this inevitably affects the cross section. Let us discuss in detail the
deeply inelastic scattering example in Fig.\ \ref{PartonPicture} (a). The
incoming lepton fluctuates into a lepton and a photon, $e \to e' \gamma^\ast$,
and the latter interacts with partons in the target $| N \rangle$
fragmenting into a number of hadrons in the final state $| n \rangle$. Since
the electromagnetism is very weak as compared to the strong interaction, one
can restrict the analysis to a single-photon exchange with spacelike momentum
$q$, $q^2 \equiv - Q^2$. In order to initiate a hard scattering, the electron
has to pass close to one of the partons, i.e., at the distance $z^2 \sim 1/Q^2$,
to exchange a photon of virtuality $Q^2$. This process is described by the
amplitude $\langle n | j_\mu | N \rangle$ with the local quark electromagnetic
current $j_\mu$. The measurement is totally inclusive with respect to final
states and only the scattered lepton is detected. The cross section reads
\begin{equation}
\label{DIS-Compton}
\sigma \left( x_{\rm B}, Q^2 \right)
=
\frac{1}{\pi} \Im{\rm m}
\ i \! \int d^4 z \, {\rm e}^{i q \cdot z}
\langle N(p) |
T \left\{
j_\mu^\dagger (z) j_\mu (0)
\right\}
| N(p) \rangle
\, .
\end{equation}
Let us turn to the physical picture of the deep-inelastic event in the
photon-hadron center-of-mass frame. For a highly virtual photon, the transverse
distance probed by it, in a Lorentz contracted hadron, is of order
$\delta z_\perp \sim 1/Q$. Due to a Lorentz dilation, the virtual photon
`sees' the partons in a frozen state during the time of transiting the target.
Therefore, since the number of parton which carry the bulk of the hadron
momentum is small, the photon will `see' mostly only one parton per collision.
The probability for multiparton correlations is suppressed by the hard momentum,
\begin{equation}
n \mbox{-parton probability}
\sim
\left(
\left( \delta z_\perp \right)^2/{\mit\Sigma}_\perp
\right)^n
\sim
1/\left( Q^2 {\mit\Sigma}_\perp \right)^n
\, ,
\end{equation}
where ${\mit\Sigma}_\perp = \pi R_N^2$ is the transverse area of the nucleon
of the radius $R_N$. These power suppressed multiparticle correlations go under
the name of higher twists. Thus, to leading approximation one can restrict
considerations to a single-parton scattering at high $Q^2$. As compared to
the inclusive annihilation mentioned above, the underlying physical picture
for the forward Compton scattering admits a classical trajectory. A quark
taken from the hadron absorbs the virtual photon and, as a result, accelerates.
Then it reemits a photon and falls in the same momentum state. After the
energy is freed into the final state the parton merges back into the parent
hadron. As we already discussed above, the process is not infrared safe and
depends on the quark binding inside the nucleon. The higher is the virtuality
of the virtual photon $Q^2$, the shorter are the distances traveled by the
parton. The points of absorption and emission are separated by a light-like
distance. The character of relevant distances in the Compton amplitude is a
consequence of deep Euclidean kinematics $Q^2 \to \infty$. Large $Q^2$ and
energies $\nu \equiv p \cdot q$, at fixed $x_{\rm B} = Q^2/(2 \nu)$, probe
short distance and time structures of the process, respectively. To identify
scales involved in deeply inelastic scattering let us go to a reference
frame where the target proton is at rest and the virtual photon's
three-momentum is directed along the $z$-axis, $q = (q^0, 0, 0, q^3)$. Then
\begin{equation}
q
=
\left(
\frac{Q^2}{2 M x_{\rm B}} ,
0,
0,
\frac{Q^2}{2 M x_{\rm B}} \sqrt{1 + 4 M^2 x^2_{\rm B}/Q^2}
\
\right) \, .
\end{equation}
For light-cone components of the momentum transfer, $q_{\pm} \equiv (q_0 \pm q_3)
/\sqrt{2}$, we have for a large momentum transfer $Q^2$
\begin{equation}
q_- \sim Q^2 / \left( M x_{\rm B} \right)
\, , \qquad
q_+ \sim M x_{\rm B}
\, .
\end{equation}
The integrand in Eq.\ (\ref{DIS-Compton}) is an oscillatory function and thus
gives vanishing result unless the distances involved are
\begin{equation}
z_- \sim 1 / \left( M x_{\rm B} \right)
\, , \qquad
z_+ \sim M x_{\rm B}/ Q^2
\, .
\end{equation}
Therefore, provided transverse separations $z_\perp$ are small, the deeply
inelastic scattering probes strong interaction dynamics close to the light-cone
$z^2 \approx 0$, and we can neglect the dependence on all coordinate components
except for $z_-$. The latter is called Ioffe time and has the meaning of the
longitudinal distance probed in the process. Its Fourier conjugated variable is
the momentum fraction $x$ of the nucleon carried by a parton interacting with
the probe. In lowest order approximation $x = x_{\rm B}$.

\begin{figure}[t]
\unitlength1mm
\begin{center}
\mbox{
\begin{picture}(0,32)(80,0)
\put(44,0){\insertfig{7}{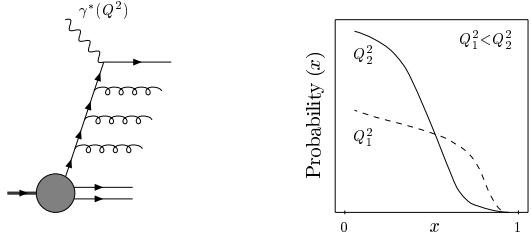}}
\end{picture}
}
\end{center}
\caption{\label{DisScalingViolation} QCD amplitude for deeply inelastic
scattering from a quark in the nucleon and resulting scaling violation
(evolution) of the probability to find a parton the with a given momentum
fraction $x$ of the parent hadron with a change of the virtuality of
$\gamma^\ast$.}
\end{figure}

A struck quark can emit a gluon or many gluons before actual interaction with
the photon, see the left drawing in Fig.\ \ref{DisScalingViolation}. When it
spills a gluon the scattered quark looses a portion of its original momentum.
Although it can be small for a single event, quarks tend to emit more being a
relativistically moving particle. Thus, the bremsstrahlung of many gluons
drifts the quark into the low momentum fraction region. Since gluons in turn
can decay into a quark-antiquark pair, there is a proliferation of partons with
small momentum fractions and correspondingly a decrease in the large-$x$ domain.
Thus, the probability to find a quark with small momentum fraction is higher at
larger $Q^2$, see the right drawing in Fig.\ \ref{DisScalingViolation}. The same
phenomenon can be viewed from the resolution perspective discussed above. The
higher $Q^2$ the smaller is the distance probed by the photon. Therefore, one
sees more and more partons in the cloud forming the parent `fat' quark.

Since the hard subprocess occupies a very small space-time volume, the scales
involved in the formation of their nonperturbative wave function are much
larger, of order of a typical hadronic scale, $1\ {\rm GeV}$. Hence, it is
quite likely that they are uncorrelated and will not interfere. Thus, although
the process depends on the hadronic state the parton has come from, this is
basically irrelevant for the hard interactions occurred. Moreover, all final state
interactions cancel in the deeply inelastic process. This is exhibited by the
relation (\ref{DIS-Compton}) stemming form the optical theorem. Thus, there is no
sensitivity to soft final-state interactions. The quantum mechanical incoherence
property of physics at different scales results into the factorization property
of the cross section (\ref{DIS-Compton}). This is the base for the predictive
power of perturbative QCD in many hadronic processes which one writes as
\begin{equation}
\label{Factorization}
\sigma \left( x_{\rm B}, Q^2 \right)
=
\int d x \ C \left( x_{\rm B}/ x; Q^2 / \mu^2 \right) \, F \left( x ; \mu^2 \right)
\, ,
\end{equation}
where $F$ is a parton distribution, --- the probability to find a parton in
the nucleon with the momentum fraction $x$, --- and $C$ is a perturbatively
computable short-distance coefficient function.

\section{Evolution equations}

Due to intrinsic ultraviolet divergences, parton distributions and coefficients
functions in Eq.\ (\ref{Factorization}) depend on a renormalization scale $\mu^2$.
However, it cancels, as it has to, in the total result since the left-hand side
being a physical observable does not depend on the arbitrary scale $\mu^2$. Thus,
factorization implies evolution as a consequence of this independence,
\begin{equation}
\frac{d \sigma}{d \ln \mu^2} = 0
\, .
\end{equation}
Separation of variables results into complementarity equations for $F$ and $C$
\begin{eqnarray}
\frac{d}{d \ln \mu^2}
C (x; Q^2 / \mu^2)
&=&
- \int d x' \, P (x/x' ; \alpha_s) C (x'; Q^2 / \mu^2)
\, , \nonumber\\
\frac{d}{d \ln \mu^2}
\label{DGLAPee}
F (x; \mu^2)
&=&
\int d x' \, P (x/x' ; \alpha_s) F (x'; \mu^2)
\, .
\end{eqnarray}
The coefficient function $C$ is computable perturbatively and so is the evolution
kernel $P$. The second of these equation is an inclusive evolution equation and
allows to predict $F$ at a scale $\mu^2$ from the function $F$ determined at
some reference scale $\mu_0^2$. Obviously, a change in $\mu^2$ is in one-to-one
correspondence to a change in $Q^2$ \footnote{We will not discuss presently the
evolution in longitudinal momentum with fixed transverse scales. This domain
is described by another type of evolution equations addressed in \cite{Kov02}.}.

\begin{figure}[t]
\unitlength1mm
\begin{center}
\mbox{
\begin{picture}(0,70)(55,0)
\put(0,10){\insertfig{5}{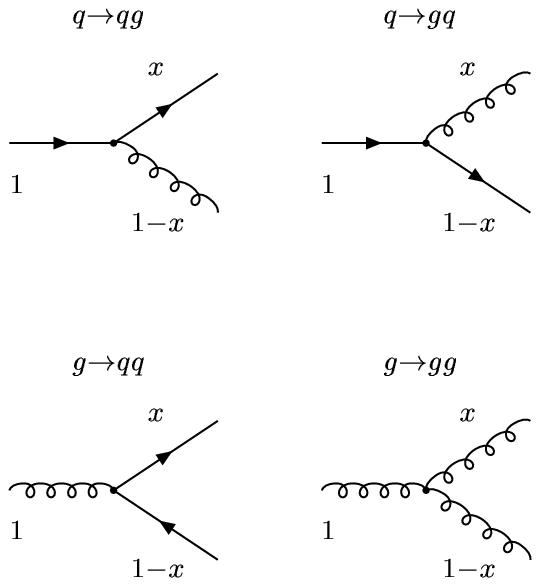}}
\put(60,0){\insertfig{5}{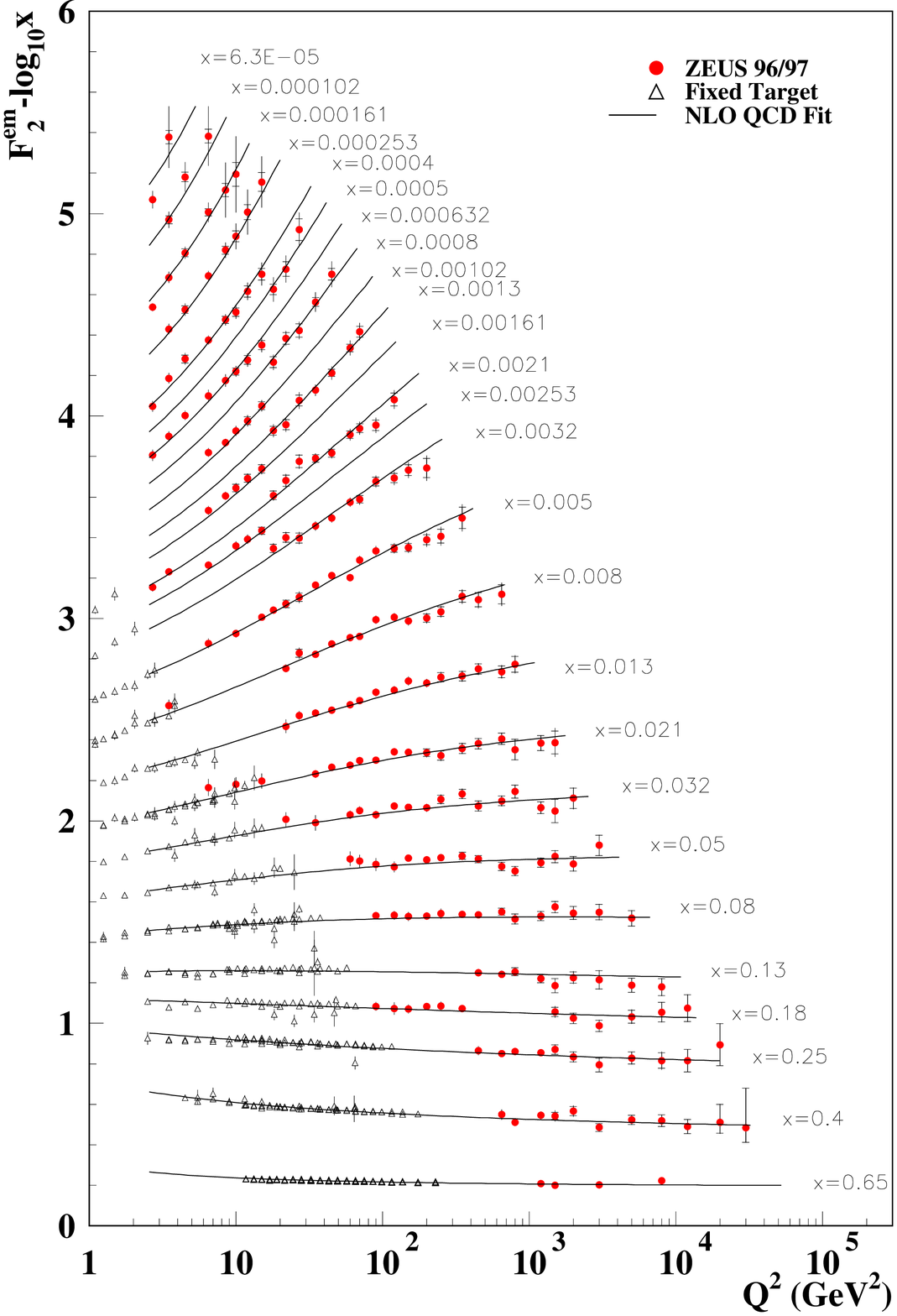}}
\end{picture}
}
\end{center}
\caption{\label{ExperimentalData} Elementary $a \to bc$ parton splitting
amplitudes encountered at each step of the evolution (left) and a fit to
experimental data \protect\cite{Zeus01} (right) implementing them (dressed with
next-to-leading order perturbative corrections) in the evolution equation
(\ref{DGLAPee}).}
\end{figure}

In the previous discussion we talked only about quark momentum probabilities.
To describe a real experiment one has to account for another type of partons
which are equally important dynamical degrees of freedom --- gluons. Gluons
contribute on equal footing with quarks and both enter in a multiplet,
$(F_q , F_g)$. This multiplet obeys a matrix evolution equation of the type
(\ref{DGLAPee}) with a quark-gluon mixing matrix of kernels
\begin{equation}
\label{SplittingKernels}
\mbox{\boldmath$P$}
=
\left(
\begin{array}{cc}
{^{qq}\!P} & {^{qg}\!P} \\
{^{gq}\!P} & {^{gg}\!P}
\end{array}
\right)
\, , \qquad \mbox{with}\qquad
{^{ab}\!P} (x) = | A_{a \to bc} |^2
\, ,
\end{equation}
where $A_{a \to bc}$ represent amplitudes of elementary splittings, demonstrated
in Fig.\ \ref{ExperimentalData}. The solution to the evolution equation (\ref{DGLAPee}),
where one includes also QCD radiative corrections as well, results to a perfect
description of experimental data for the unpolarized deeply inelastic cross section,
shown in the same figure.

We consider below a more general case of functions which interpolates between
inclusive parton distribution and exclusive distributions amplitudes and, thus,
possess a very rich structure. They are known as generalized parton distributions
\cite{MulRobGeyDitHor94,Ji96,Rad96} $F (x, \eta)$ and are measurable in a number
of hard exclusive reactions \cite{Ji98,Rad01,GoePolVan01,BelMulKir01} such as in
Fig.\ \ref{PartonPicture} (b) with a photon emission into the final state. They
admit an operator representation which reads, e.g., for quarks
\begin{equation}
\langle N(p_2) | \bar q ( - z_-) q (z_-)| N(p_1) \rangle
=
\int dx \, {\rm e}^{- i x z_- (p_1 + p_2)_+} F (x, \eta)
\, .
\end{equation}
It is a correlation function of fields on the light cone separated by the
distance $2 z_-$ which acquire a scale dependence under renormalization.
We omitted an inessential to our present consideration Dirac matrix
projecting quark spinor indices. Due to different momenta of incoming and
outgoing hadrons, the specifics of these functions lies in a presence of a
non-zero $t$-channel momentum transfer $\eta = (p_2 - p_1)_+/2$, varying which
one probes correlations of partons in different momentum states. The evolution
equation has the same schematic form as (\ref{DGLAPee}),
\begin{equation}
\frac{d}{d \ln \mu^2}
F (x, \eta; \mu^2)
=
\int d x' \, K (x/\eta , x'/\eta ; \alpha_s) F (x', \eta; \mu^2)
\, .
\end{equation}
The evolution kernel interpolates between two known limits, i.e.,
\begin{eqnarray*}
\lim_{\eta \to 0} K (x/\eta , x'/\eta ; \alpha_s)
= P (x/x'; \alpha_s)
\, , \quad
\lim_{\eta \to 1} K (x/\eta , x'/\eta ; \alpha_s)
= V (x, x'; \alpha_s)
\, ,
\end{eqnarray*}
the fully inclusive case we have discussed above $(\eta = 0)$ and another one
which corresponds to distribution amplitudes which enter the description of a
number of exclusive processes \cite{BroLep80,EfrRad80}.

The moments of the function $F (x, \eta)$ correspond to the operator
matrix elements of local operators, e.g., for quarks,
\begin{equation}
\label{OperatorsWithout}
\int dx \, x^j F (x, \eta)
=
\langle P_2 |
\bar q \!
\stackrel{\leftrightarrow}{\partial}{\!}^j_+
q
| P_1 \rangle
\, ,
\end{equation}
where $\stackrel{\leftrightarrow}{\partial} \equiv
\stackrel{\rightarrow}{\partial} \! - \! \stackrel{\leftarrow}{\partial}$. Under
renormalization the operators in Eq.\ (\ref{OperatorsWithout}) get mixed with
operators containing total derivatives $\stackrel{\phantom{\leftarrow}}{\partial}
\equiv \stackrel{\rightarrow}{\partial} \! + \! \stackrel{\leftarrow}{\partial}$
but, indeed, the same total number of derivatives as a consequence of Poincar\'e
invariance
\begin{equation}
\label{LocalOperators}
{\cal R}_{jk} =
\bar q \! \stackrel{\phantom{\leftrightarrow}}{\partial}{\!}^k_+ \!
\stackrel{\leftrightarrow}{\partial}{\!}^{j - k}_+ q
\, .
\end{equation}
Since the operators with total derivatives matter for each moment we have to
diagonalize a $j \times j$ matrix equation,
\begin{equation}
\label{Mixing}
\frac{d}{ d \ln \mu^2}
\left(
\begin{array}{c}
{\cal R}_{j0} \\
\vdots \\
{\cal R}_{jj}
\end{array}
\right)
=
- \frac{1}{2}
\left(
\begin{array}{ccc}
\gamma_{jj} & \cdots & \gamma_{j0} \\
            & \ddots & \vdots \\
            &        & \gamma_{00}
\end{array}
\right)
\left(
\begin{array}{c}
{\cal R}_{j0} \\
\vdots \\
{\cal R}_{jj}
\end{array}
\right) \, ,
\end{equation}
with zeros below the diagonal. The matrix has obviously a triangular
form since the operator containing at least one total derivative cannot
mix with operators involving none of them. Otherwise one would run in a
contradiction for the forward matrix elements where total derivatives
are irrelevant $\langle p | \partial ( \cdots )| p \rangle = 0$.
Another important observation is that the diagonal elements of the
anomalous dimensions are the moments of the splitting functions
(\ref{SplittingKernels})
\begin{equation}
{^{ab}\gamma^{\rm D}_j} = - 2 \int dx \, x^j \, {^{ab}\!P} (x) \, .
\end{equation}

\section{Use of conformal symmetry}

Recall the diagonalization of the Schr\"odinger equation for a hydrogen atom.
The $O (3)$ symmetry of the potential allows to reduce the three-dimensional
problem to a one-dimensional one, i.e., one performs a partial wave decomposition
in terms of spherical harmonics. The spherical harmonics are eigenfunctions of
the quadratic Casimir operator of the rotation group.

For the case at hand one can use the space-time symmetries of QCD Lagrangian
to simplify significantly and ultimately solve the problem. Classical QCD is
Poincar\'e, i.e., Lorentz ${\cal M}_{\mu\nu}$ and translation ${\cal P}_\mu$,
invariant as any realistic field theory describing nature, a well as due to
absence of any dimensionful parameters for massless quarks, invariant under
dilatation ${\cal D}$ and special conformal ${\cal K}_\mu$ transformations
\cite{MacSal69}. Coordinates transform as follows under them,
\begin{eqnarray*}
z_\mu \stackrel{{\cal M}}{\to}
\omega_{\mu\nu} z_\nu
\, , \quad \
z_\mu \stackrel{{\cal P}}{\to}
z_\mu + a_\mu
\, , \quad \
z_\mu \stackrel{{\cal D}}{\to}
\lambda z_\mu
\, , \quad \
z_\mu \stackrel{{\cal K}}{\to}
c_\nu \left( z^2 g_{\mu\nu} - z_\mu z_\nu \right)
\, .
\end{eqnarray*}
The special conformal transformation can be visualized as a sequence
${\cal K}_\mu \equiv {\cal I} {\cal P}_\mu {\cal I}$ of inversion
${\cal I} z_\mu = z_\mu / z^2$, translation and inversion. This fifteen
transformations form a group, $SO (4,2)$. It is easy to construct the
representation of the group on field operators using induced
representations \cite{MacSal69}. For instance, $\delta_\mu^{\cal P} q
\equiv i[ q, {\cal P}_\mu]_- = - \partial_\mu q$, etc. The eigenvectors
of the quadratic Casimir operator of the projected collinear conformal group,
\begin{equation}
\mbox{\boldmath$J$}{}^2
= \frac{1}{2} {\cal P}_+ {\cal K}_-
-
\frac{1}{4} \left( {\cal D} + {\cal M}_{-+} + 2 i \right)
\left( {\cal D} + {\cal M}_{-+} \right) \, ,
\end{equation}
in the basis of the bilinear operators (\ref{LocalOperators}) are called
conformal operators with eigenvalues\footnote{The indices attached to
$\mbox{\boldmath$J$}{}^2$ stand for the conformal spin $j_a = (d_a + s_a)/2$
of fields, having the dimension $d_a$ and spin $s_a$, which build ${\cal O}_{jl}$.}
\begin{equation}
\mbox{\boldmath$J$}{}^2_{a b} {\cal O}_{jl}
= J_{ab} (J_{ab} - 1 ) {\cal O}_{jl}
\, , \qquad
J_{ab} = j + j_a + j_b \, .
\end{equation}
For quarks $j_q = 1$ and conformal operators are given in terms of Gegenabauer
polynomials \cite{BroLep80,EfrRad80} ($l \geq j$)
\begin{equation}
{\cal O}_{jl} =
\bar q \stackrel{\phantom{\leftrightarrow}}{\partial}{\!}^l_+ C_j^{2 j_q - 1/2}
\left(
\stackrel{\leftrightarrow}{\partial}_+ \!\!
/
\!\! \stackrel{\phantom{\leftrightarrow}}{\partial}_+
\right) q
\, ,
\end{equation}
which is a very specific combination of local operators ${\cal R}_{jk}$
(\ref{LocalOperators}).

Generally, the renormalization group equation reads for the conformal operators
\begin{equation}
\frac{d}{ d \ln \mu^2}
{\cal O}_{jl}
= - \frac{1}{2} \sum_{k = 0}^j
\left(
\gamma^{\rm D}_{j} \delta_{jk} + \gamma^{\rm ND}_{jk}
\right)
{\cal O}_{kl}
\, .
\end{equation}
As is well known a renormalization group equation is a Callan-Symanzik equation,
i.e., the scale Ward identity for a Green function with a (conformal, for the
case at hand) operator insertion,
\begin{equation}
\label{Green}
{\cal G} (z_1, \dots, z_N)
\equiv
\int D q D \bar q D A {\rm e}^{i \int d z \, {\cal L} (z)}
\, {\cal O}_{jl} \, \bar q (z_1) \dots q (z_k) \dots A (z_N) \, .
\end{equation}
The anomalous dimensions of operators arise due to development of an anomaly
in the trace of the energy-momentum tensor,
\begin{equation}
\delta^{\cal D} {\cal L} = - {\mit\Theta}_{\mu\mu}
=
\frac{\beta_d}{g} \left( G^a_{\mu\nu} \right)^2 \, ,
\end{equation}
where $\beta_d$ is a $d$-dimensional QCD beta function $\beta_d = \beta +
g (d - 4)/2$. We did not display other terms which could not affect the
physical sector \cite{BelMul98}. The anomalous dimensions show up in the
renormalization of an operator product,
\begin{equation}
\label{ScaleAnomaly}
{\cal O}_{jl} \, i \! \int \! d^d z \, {\mit\Theta}_{\mu\mu} (z)
=
\sum_{k = 0}^j \left(
\gamma^{\rm D}_{j} \delta_{jk} + \gamma^{\rm ND}_{jk}
\right)
{\cal O}_{kl}
\, .
\end{equation}

Since the conformal algebra is fulfilled on interacting (Heisenberg) field
operators one can derive a constraint on the anomalous dimension from it.
It is provided by the commutator
\begin{equation}
\left[ {\cal D}, {\cal K}_\mu \right] = i {\cal K}_\mu \, ,
\end{equation}
which results into a commutator equation \cite{Mul94,BelMul98}
\begin{equation}
\label{Constraint}
\left[
\gamma^{\rm D} + {\gamma}^{\rm ND}
,
a + \gamma^c + 2 \frac{\beta}{g} b \,
\right]_-
= 0
\, ,
\end{equation}
in terms of $\alpha_s$-independent matrices $a$ and $b$; QCD beta-function
$\beta$ and a special conformal anomaly $\gamma^c$ which arises, similarly to
Eq.\ (\ref{ScaleAnomaly}), in the renormalization of the product
\begin{equation}
{\cal O}_{jl} \int \! d^d z \, \delta^{\cal K}_- {\cal L} (z)
=
\sum_{k = 0}^j \gamma^c_{jk} {\cal O}_{k, l - 1} \, .
\end{equation}
The special conformal variation of the action is given in terms of
a coordinate moment of the energy-momentum tensor, $\delta^{\cal K}_-
{\cal L} (z) = - 2 z_- {\mit\Theta}_{\mu\mu} (z)$.

One can immediately draw important conclusions from Eq.\ (\ref{Constraint}).
If one is interested in the one-loop anomalous dimension matrix, ${\cal O}
(\alpha_s)$, one has to keep the $\alpha_s$-independent term $a$ only and
write $\left[ \gamma^{\rm D (0)} + \gamma^{\rm ND (0)}, a \right] = 0$. Since
the matrix $a$ is diagonal \cite{BelMul98} we conclude that $\gamma^{\rm ND (0)}
= 0 \cdot \alpha_s$. At next-to-leading order, we have to account for $\beta$
and $\gamma^c$ in one-loop approximation. This provides via Eq.\ (\ref{Constraint})
an expression for two-loop nondiagonal elements $\gamma^{\rm ND (1)}$ in terms
of $\gamma^{\rm D (0)}$, $\beta_0$ and $\gamma^{c (0)}$,
\begin{equation}
\gamma^{\rm ND (1)} =
\left[
\gamma^{\rm D (0)}
,
\frac{b}{a} \left( \beta_0 - \gamma^{\rm D (0)} \right)
+
\frac{\gamma^{c (0)}}{a}
\right]_-
\, .
\end{equation}
No two-loop calculation has to be done! The diagonal anomalous dimensions
$\gamma^{\rm D}$ coincide with the anomalous dimension of local operators
without total derivatives and are available in the literature on deeply
inelastic scattering.

\section{Use of superconformal symmetry}

As we shall see momentarily, supersymmetry provides relations between the
anomalous dimensions of conformal operators, while the extension to
superconformal case results into identities between the special conformal
anomalies $\gamma^c$ of quark and gluon operators.

If one puts the QCD quarks into the adjoint representation of the color
group and sets the number of flavors to one, one recovers ${\cal N} = 1$
super-Yang-Mills theory in the Wess-Zumino gauge. In four-dimensional
space-time the Lagrangian is invariant under superconformal group, i.e.,
$SO(4,2)$ discussed above plus translational ${\cal Q}$ and superconformal
${\cal S}$ transformations \cite{WesZum74a,Fer74,DonSon74},
\begin{equation}
\label{SuperTrans}
\delta^{\cal F} q^a
=
\frac{i}{2} G^a_{\mu\nu} \sigma_{\mu\nu} \xi - i D^a \gamma_5 \xi
\, , \
\delta^{\cal F} B^a_\mu
=
- i \bar \xi \gamma_\mu q^a
\, , \
\delta^{\cal F} D^a
=
\bar \xi {\cal D}^{ab}_\mu \gamma_\mu \gamma_5 q^a
\, ,
\end{equation}
with non-dynamical auxiliary field $D^a$ and transformation parameter $\xi
\equiv \xi_0 - i \! \not \! x \xi_1$. The parameter $\xi_0$ ($\xi_1$)
parametrizes ${\cal F} = {\cal Q}$ (${\cal F} = {\cal S}$) transformations.

One can construct an ${\cal N} = 1$ chiral superfield from certain linear
combinations of bosonic, i.e., $\bar qq$ and $gg$, and fermionic, i.e.,
$gq$, conformal operators, see Refs.\ \cite{BukFroKurLip85,BelMulSch98}. So
that under (\ref{SuperTrans}) they transform as Wess-Zumino multiplet. This
covariance simplifies the derivation of relations between the anomalies.

We follow the same procedure as in the previous section by using the commutator
algebra of the supeconformal group for the Green function ${\cal G} (z_1, \dots,
z_N)$ (\ref{Green}) and use Ward identities to find out consequences. Because
of the quantization which require a regularization procedure, the variation of
the action develops an anomaly in $d$-dimensions for the ${\cal S}$-transformation
but not for the ${\cal Q}$-transformation
\begin{equation}
\label{SusyVariation}
\delta^{\cal Q} {\cal L}
= 0
\, , \qquad
\delta^{\cal S} {\cal L}
=
\frac{4 - d}{2} \bar \xi_1 i \sigma_{\mu\nu} G^a_{\mu\nu} q^a
\, .
\end{equation}
Although there arise contributions in both due to explicit supersymmetry
breaking by gauge fixing, however, these sources cannot affect gauge-invariant
quantities we deal with. For our purposes we need two commutators from the
algebra
\begin{equation}
\label{SuperConfAlg}
\left[ {\cal Q} , {\cal D} \right]_- = \frac{i}{2} {\cal Q}
\, , \qquad
\left[ {\cal Q} , {\cal K}_\mu \right]_- = \gamma_\mu {\cal S}
\, .
\end{equation}

In view of Eq.\ (\ref{SusyVariation}) the first one is non-anomalous and
results into a number of relations between the elements of the quark-gluon
mixing matrix \cite{BukFroKurLip85,BelMulSch98,BelMul99}, with the most
trivial example involving only the diagonal elements
\begin{equation}
\label{Dokshitzer}
{^{qq}\gamma^{\rm D}_j} + \frac{6}{j} \, {^{gq}\gamma^{\rm D}_j}
= \frac{j}{6} \, {^{qg}\gamma^{\rm D}_j} + {^{gg}\gamma^{\rm D}_j}
\, .
\end{equation}
The second commutator in Eq.\ (\ref{SuperConfAlg}) possesses the superconformal
anomaly (\ref{SusyVariation}) and, as a result, relations of the type
(\ref{Dokshitzer}), but with $\gamma$ being replaced by $\gamma^c$, are
affected in a controllable manner by a computable addendum $\Delta$ which
arises from renormalization of the operator product \cite{BelMul00b}
\begin{equation}
{\cal O}_{jl} \, i \! \int \! d^d z \, \delta^S {\cal L} (z)
=
\sum_{k = 0}^j \Delta_{jk} {\cal O}_{k, l - 1} \, .
\end{equation}

The knowledge of a complete set of such relations allows to compute all
entries of the quark-gluon mixing matrix from the knowledge of only one
element, e.g., ${^{qq}\gamma^{\rm D}} + {^{qq}\gamma^{\rm ND}}$. This
procedure has successfully been applied in the reconstruction of two-loop
off-forward evolution kernels \cite{BelFreMul99}.

\section{Integrability of three-particle problem}

\begin{figure}[t]
\unitlength1mm
\begin{center}
\mbox{
\begin{picture}(0,17)(38,0)
\put(0,0){\insertfig{7}{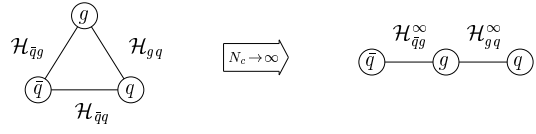}}
\end{picture}
}
\end{center}
\caption{\label{LargeNcReduction} Large-$N_c$ reduction of three-particle
$\bar q g q$ Hamiltonian.}
\end{figure}

As we explained in the introduction, multi-particle correlations arise as
power suppressed effects in hard cross sections and exhibit a genuine quantum
mechanical interference of hadron wave functions with different number of
constituents. The first non-trivial and the most interesting example is
three-particle functions which emerge in hard forward \cite{JafJi92} and
off-forward \cite{BelMul00a} Compton scattering in the form of quark-gluon-quark
$\langle \bar q g q \rangle$ and three-gluon $\langle g g g \rangle$ correlations.
Another prominent example is baryon distribution amplitudes \cite{BroLep80}.
For the sake of definiteness, let us discuss quark-gluon-quark correlators.
As all other twist-three sectors, they are distinguished from all even
higher-particle sectors by the fact that their basis can be reduced to the
one consisting only of quasi-partonic operators \cite{BukFroKurLip85}. This
implies that the three-particle kernel of the leading order evolution equation
in QCD coupling constant,
\begin{equation}
\frac{d}{d \ln \mu^2}
F \left( \{x\}; \mu^2 \right)
= \frac{\alpha_s}{2 \pi} \int \! \prod^3_i d x_i \,
\delta \left( \sum^3_j x_j - \eta \right)
K \left( \{ x \}; \{ x' \}\right)
F \left( \{x'\}; \mu^2 \right) ,
\end{equation}
with $\{ x \} = x_1, x_2, x_3$, is merely given by the sum of pair-wise
interactions of twist-two kernels discussed in the previous sections,
$K = K_{\bar q g} + K_{q g} + K_{\bar q q}$. As a result of the
conformal covariance, established there, the pair-wise kernels in the
basis of local conformal operators, to be called ${\cal H}_{ab}$ later,
depend only on two-particle Casimir operators, $\mbox{\boldmath$J$}{}^2_{ab}
\equiv J_{ab} (J_{ab} - 1)$, of the conformal group,
\begin{equation}
{\cal H}
= {\cal H}_{\bar q g} \left( J_{\bar q g} \right)
+ {\cal H}_{q g} \left( J_{q g} \right)
+ {\cal H}_{\bar q q} \left( J_{\bar q q} \right)
\, .
\end{equation}
The color structure of pair-wise interaction allows to simplify further
the diagonalization problem by noticing that ${\cal H}_{\bar q q}$
vanishes as $1/N_c^2$ relative to ${\cal H}_{(\bar q, q) g}$ in multicolor
limit, i.e., $N_c \to \infty$. Therefore, the three-particle problem on a
ring is reduced to the one on a line with the interaction of end sites being
neglected, see Fig.\ \ref{LargeNcReduction}. Our cherished goal is to solve a
Schr\"odinger equation
\begin{equation}
{\cal H}^\infty {\mit\Psi}
\equiv
\left( {\cal H}^\infty_{\bar q g} + {\cal H}^\infty_{q g} \right) {\mit\Psi}
=
N_c {\cal E}^\infty {\mit\Psi}
\, .
\end{equation}

The explicit form of the pair-wise Hamiltonians depends on the $t$-channel
quantum numbers. For instance, for the operators of the Dirac-color structure
$\bar q \sigma_{\rho \{ \mu} t^c G^c{\!\!}_{\nu \} \rho} q$ they are
\begin{equation}
\frac{1}{N_c} {\cal H}^{\infty}_{ab} (J_{ab})
= \psi \left( J_{ab} + 3/2 \right)
+ \psi \left( J_{ab} - 3/2 \right)
- 2 \psi (1) \, ,
\end{equation}
where $\psi (x) = d \ln {\mit\Gamma} (x) / dx$ and the indices run over
$a = (\bar q, q)$ and $b = g$. The large-$N_c$ Hamiltonian ${\cal H}^{\infty}$
describes a three-site inhomogeneous integrable open spin chain model. Integrability
means that it possesses a complete set of integrals of motion matching the number
of degrees of freedom. The problem admits a representation in terms of the
${\cal R}$-matrix acting on a tensor product of vector spaces $V_a \otimes V_b$
and satisfying Yang-Baxter relation \cite{Fad96},
\begin{equation}
\label{YBequation}
\hspace*{-3cm}
{\cal R}_{ab} (\lambda)
\equiv
{
\begin{picture}(0,0)(0,0)
\put(0,-6){\insertfig{1}{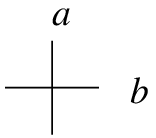}}
\end{picture}
}
\hspace{1cm}
\, ,
\phantom{\frac{\mbox{\large A}}{\mbox{\large A}}A}
{
\begin{picture}(0,0)(0,0)
\put(0,-11){\insertfig{3}{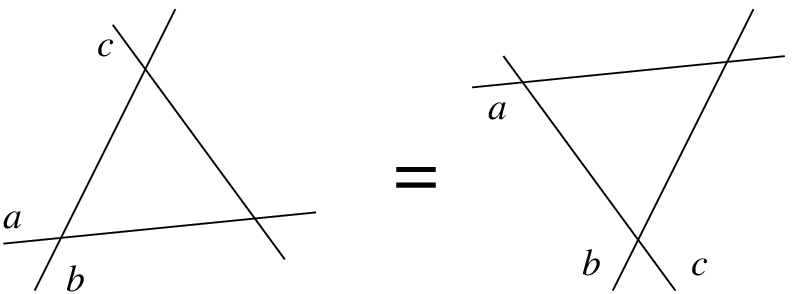}}
\end{picture}
}
\phantom{\mbox{\Huge A}}
\end{equation}
One of the solutions to the Yang-Baxter equation is given by ${\cal R}_{ab}
(\lambda) \sim {\mit\Gamma} (J_{ab} + \lambda) / {\mit\Gamma} (J_{ab} - \lambda)$.
The construction by Sklyanin \cite{Skl88} allows to find the Hamiltonian
corresponding to this ${\cal R}$-matrix via the following graphic equation
\cite{Bel99,DerKorMan99}
\begin{equation}
\label{HamiltonFromYB}
{\cal H}^\infty = \frac{1}{2}
\left.
\frac{d}{d \lambda}
\right|_{\lambda = 0}
\ln
\Bigg\{
{
\begin{picture}(0,0)(0,0)
\put(0,-13.1){\insertfig{2.3}{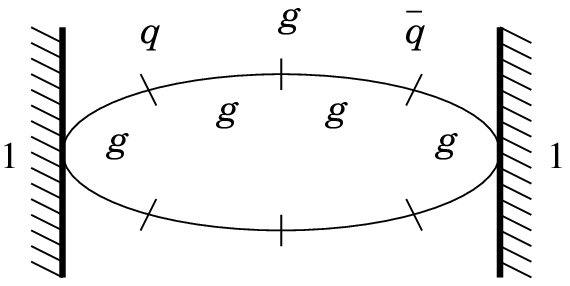}}
\end{picture}
}
\hspace{2.3cm}
\Bigg\} \, .
\end{equation}
The spectral parameter $\lambda$ in the $q$ and $\bar q$ sites
is shifted as $\lambda \pm 3/2$. The picture is self explanatory: the
three ${\cal R}$-matrices with dimensions of spaces as displayed are
aligned next to each other, then reflected with the unity matrix and
traced in the auxiliary space with their inverse. The resulting object
is called the transfer matrix. Its logarithmic derivative determines
the $qgq$ Hamiltonian up to an additive constant.

The Yang-Baxter equation holds for an arbitrary dimension of the vector spaces.
When the auxiliary space is two-dimensional, one finds another solution to
(\ref{YBequation}), ${\cal R}_{a,1/2} (\lambda) = \lambda 1 + \sigma^i J^i_a$.
Calculating the transfer matrix for the two-dimensional auxiliary space, one
identifies two operators \cite{BraDerMan98,Bel99,DerKorMan99}
\begin{equation}
\mbox{\boldmath$J$}{}^2
= \mbox{\boldmath$J$}{}^2_{\bar q g}
+ \mbox{\boldmath$J$}{}^2_{q g}
+ \mbox{\boldmath$J$}{}^2_{\bar q q}
\, , \qquad
\mbox{\boldmath${\cal Q}$}
= \left[
\mbox{\boldmath$J$}{}^2_{q g} , \mbox{\boldmath$J$}{}^2_{\bar q g}
\right]_+
- 2 (3/2)^2
\left(
\mbox{\boldmath$J$}{}^2_{q g}
+
\mbox{\boldmath$J$}{}^2_{\bar q g}
\right)
\, ,
\end{equation}
which commute with the Hamiltonian (\ref{HamiltonFromYB}) by construction
since all of them are deduced from bundles satisfying the Yang-Baxter
equation.

Thus instead of diagonalizing the Hamiltonian one can attempt to solve a
much simple eigenvalue problem for the fourth-order differential operator
$\mbox{\boldmath${\cal Q}$}$. Because of commutativity, the Hamiltonian
shares the same eigenfunctions ${\mit\Psi}$. Thus the eigenvalues
${\cal E}^\infty$ of ${\cal H}^\infty$ are functions of quantized values
$J$ and $Q$ of the integrals of motion $\mbox{\boldmath$J$}{}^2$ and
$\mbox{\boldmath${\cal Q}$}$. The problem was solved using WKB, i.e.,
large-$J$, approximation with the result for energy levels
\cite{BraDerMan98,Bel99,DerKorMan99}
\begin{eqnarray}
{\cal E}^\infty_0 (J) &=& \psi (J + 3) + \psi (J + 4) - 2 \psi (1) - 1/2
\, , \\
{\cal E}^\infty (J, Q) &=& 2 \ln J
+ 2 \Re{\rm e} \,
\psi
\left(
3/2 + i \sqrt{( 2 Q / J^2 - 3)/4}
\right)
- 4 \psi (1) - 3/2
\, , \nonumber
\end{eqnarray}
where the special value ${\cal E}^\infty_0$ is separated from the rest of
the spectrum by a gap and was known for some time \cite{BalBraKoiTan96}.
Full account of this and other exactly solvable cases can be found in the
literature \cite{Bel99,DerKorMan99,BraDerKorMan99}.

\vspace{1mm}

We would like to thank the organizers of the workshop for such an interesting
meeting and INT (Seattle) for support. This work was supported by the US
Department of Energy under contract DE-FG02-93ER40762.

\end{document}